\documentclass[%
 aip,
 amsmath,amssymb,
 reprint,%
]{revtex4-1}

\usepackage{graphicx}
\usepackage{dcolumn}
\usepackage{bm}
\usepackage{amsmath,color,amssymb}
\usepackage[normalem]{ulem}
\usepackage{bm}
\usepackage{soul}
\usepackage[utf8]{inputenc}
\usepackage[T1]{fontenc}
\usepackage{mathptmx}
\usepackage{mathtools}

\definecolor{new}{rgb}{0.38,0.6,0.8}
\definecolor{old}{rgb}{1,0,0}
\newcommand{\new}[1]{{\color{new} #1}}

\draft 

\begin{document}


\title{Computer-automated tuning procedures for semiconductor quantum dot arrays}



\author{A. R. Mills}
\affiliation{Department of Physics, Princeton University, Princeton, New Jersey 08544, USA}

\author{M. M. Feldman}
\affiliation{Department of Physics, Princeton University, Princeton, New Jersey 08544, USA}

\author{C. Monical}
\affiliation{Sandia National Laboratories, Albuquerque, New Mexico 87185, USA}

\author{P. J. Lewis}
\affiliation{Sandia National Laboratories, Albuquerque, New Mexico 87185, USA}

\author{K. W. Larson}
\affiliation{Sandia National Laboratories, Albuquerque, New Mexico 87185, USA}

\author{A. M. Mounce}
\affiliation{Sandia National Laboratories, Albuquerque, New Mexico 87185, USA}

\author{J. R. Petta}
\affiliation{Department of Physics, Princeton University, Princeton, New Jersey 08544, USA}


\date{\today}

\begin{abstract}
As with any quantum computing platform, semiconductor quantum dot devices require sophisticated hardware and controls for operation.
The increasing complexity of quantum dot devices necessitates the advancement of automated control software and image recognition techniques for rapidly evaluating charge stability diagrams.
We use an image analysis toolbox developed in Python to automate the calibration of virtual gates, a process that previously involved a large amount of user intervention.
Moreover, we show that straightforward feedback protocols can be used to simultaneously tune multiple tunnel couplings in a triple quantum dot in a computer automated fashion.
Finally, we adopt the use of a `tunnel coupling lever arm' to model the interdot barrier gate response and discuss how it can be used to more rapidly tune interdot tunnel couplings to the GHz values that are compatible with exchange gates.
\end{abstract}

\pacs{}

\maketitle 

Quantum processors rely on classical hardware and controls in order to prepare, manipulate, and measure qubit states.
For this reason, it is advantageous to develop tools to automate the operation of small quantum processors and routinely tune-up single qubit and two-qubit gates to maintain high performance\cite{Lucero2010,Brown2011,Frank2017}. 
Semiconductor spin qubits are a promising platform for realizing quantum computation largely due to their potential for scaling\cite{Loss1998}.
To tune up semiconductor quantum dots for operation as spin qubits requires control over the ground state charge occupation and chemical potential of each dot, as well as the interdot tunnel couplings\cite{Petta2005}.

Following the recent progress in constructing high-fidelity single-qubit and two-qubit gate operations with electron spins\cite{Veldhorst2014,Yoneda2018,Zajac2018,Watson2018,Huang2019,Sigillito2019swap}, there are increasing efforts towards scaling to larger multi-qubit devices\cite{Sigillito2019site,Mills2019,Volk2019,Dehollain2019}.
One of the key challenges in scaling up spin qubits is developing the software tools necessary to keep pace with increasingly complex devices. To date, approaches to implementing automated control software during tune-up of semiconductor qubits include training neural networks to identify the state of a device\cite{Kalantre2019}, experimentally realizing automated control procedures for tuning double quantum dot (DQD) devices into the single-electron regime\cite{Baart2016auto}, and automatically tuning the interdot tunnel coupling in a DQD\cite{Botzem2018,vanDiepen2018,Teske2019}.

In this Letter, we use an image analysis toolbox developed at Sandia National Laboratories to accurately analyze charge stability diagrams acquired from a triple quantum dot (TQD) unit cell of a 9-dot linear array\cite{Zajac2016,Mills2019}.
Computer automated analysis of charge stability diagrams performs the inversion of the device capacitance matrix and the establishment of `virtual gates'. Virtual gates compensate for cross-capacitances in the device and allow the chemical potential of each dot in the array to be independently controlled\cite{Nowack2011,Hensgens2017,Mills2019}.
Furthermore, we use image analysis to locate interdot charge transitions and automatically perform measurements of the interdot tunnel coupling\cite{DiCarlo2004}.
Using simple feedback protocols, we demonstrate simultaneous tune-up of the interdot tunnel couplings in a Si/SiGe TQD.
Finally, we introduce a `tunnel coupling lever arm'\new{\cite{Rochette2019}} that quantifies the tunnel coupling as a function of gate voltage and aids in the automated tuning of quantum dot arrays\cite{Hensgens2017,Mills2019}.

\begin{figure}
\includegraphics{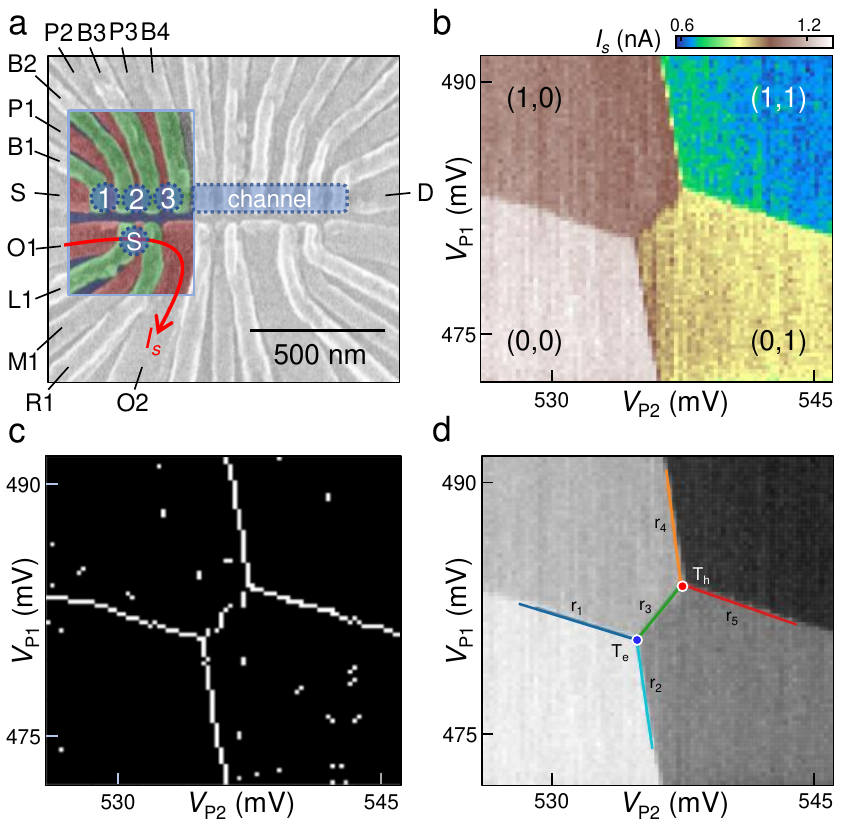}%
\caption{\label{Fig1} (a) Scanning electron microscope image of the full 12-dot device. The leftmost unit cell, consisting of dots 1, 2, and 3 in the array and the charge sensor dot `S', is false-colored. (b) DQD charge stability diagram for dots 1 and 2 as measured in the sensor dot current $I_s$. (c) Results of the edge detection algorithm with white pixels indicating likely edges. (d) The results of charge transition and triple point fitting are overlaid on the charge stability diagram.}%
\end{figure}

As illustrated in Fig.~1a, we use an overlapping gate architecture to fabricate a linear array of 9 quantum dots\cite{Zajac2015}. In the presence of a magnetic field gradient, each dot in the array can be used to define a single  ``Loss-DiVincenzo'' spin qubit\cite{Loss1998,Sigillito2019site}.
The 9-dot array is fabricated in close proximity to three quantum dot charge sensors. The device utilizes a repeating unit cell structure, with one unit cell consisting of three quantum dots and a charge sensor\cite{Zajac2016}. For the purpose of this manuscript, we operate the 9-dot device in TQD mode, with a single electron accumulated under each of the plunger gates P1, P2, and P3. A conducting channel is accumulated to the right of dot 3 and connects to a large Fermi reservoir accumulated under gate D shown to the right in Fig.~1a. 

To measure the charge occupancy of the TQD, we tune the charge sensor dot into the few electron regime with the gate voltage set to the side of a Coulomb blockade peak, where the sensor dot current $I_s$ is sensitive to changes in the local electrical potential\cite{Petta2004,Schleser2004,Vandersypen2004,Fujisawa2006,Gustavsson2006}. Figure 1b shows a DQD charge stability diagram acquired by measuring \textit{I}$_s$ as a function of the plunger gate voltages $V_{P1}$ and $V_{P2}$. Here ($N_1$,$N_2$) refers to the number of electrons in dot 1 and 2. From the charge stability diagram, we can extract information such as the cross-capacitances between a plunger gate and neighboring dots, as well as the location of the interdot charge transition in gate-voltage space. This information is usually extracted by fitting the charge transition lines by hand, which takes several minutes. Here we utilize image analysis to automatically extract this information in several seconds.

Our data analysis procedure detects an interdot charge transition [e.g. (1,0)--(0,1)] centered in a charge-stability diagram using image processing techniques implemented in Python\cite{vanRossum1995} with SciPy\cite{Jones2001}. There are five charge transitions in the vicinity of an interdot charge transition (see Fig.~1d), each of which is defined by a line segment $r_{1-5}$. Line segments $r_1$, $r_2$, and $r_3$ intersect at the electron-like triple point ($T_e$), and line segments $r_3$, $r_4$, and $r_5$ intersect at the hole-like triple point ($T_h$). We can parametrize the location of the interdot charge transition using the two triple points and four line segments, and then evaluate the parameters by image analysis techniques as detailed below\cite{Duda1972}. The visibility of the interdot charge transition can sometimes be weak in experiments due to the charge sensor placement relative to the dots, causing this transition to go undetected during automated analysis. For instance, in these experiments the charge sensor response at the (0,0) $\rightarrow$ (0,1) charge transition is approximately 4 times greater than the response at the (1,0) $\rightarrow$ (0,1) charge transition. We therefore leave the line segment $r_3$ joining the two triple points out of the parameterization. However, if $r_3$ is visible in the binary then it will enhance the fit.

The data analysis procedure to fit a local region of a charge stability diagram consists of five steps:
\begin{enumerate}
\item Convert the derivative of the charge stability diagram into a binary image of ``white'' and ``black'' pixels, where white pixels represent likely charge transitions, by thresholding the values to the 98$^{th}$ percentile. Figure~1c shows the results of the binary conversion where the white pixels largely track the five charge transitions in the data.
\item Use the Hough line transform to determine the slopes of the line segments $r_1$, $r_2$, $r_4$, and $r_5$.
\item Detect possible triple points using Hough transform-inspired accumulators in parameterized anticrossing space.\cite{Ballard1981}
\item Use template matching to select the most probable location of the interdot charge transition.
\item Perform a template-based local search to optimize the parameters of the detected interdot charge transition. 
\end{enumerate}

The results of this fitting procedure are illustrated by plotting the extracted triple points and line segments over the raw charge stability diagram, as shown in Fig.~1d. In this work we have optimized the algorithm to detect a single interdot charge transition in the charge stability diagrams. However, this algorithm is flexible enough to be optimized to find multiple interdot charge transitions in a larger-scale charge stability diagram.


\begin{figure}[t]
\includegraphics{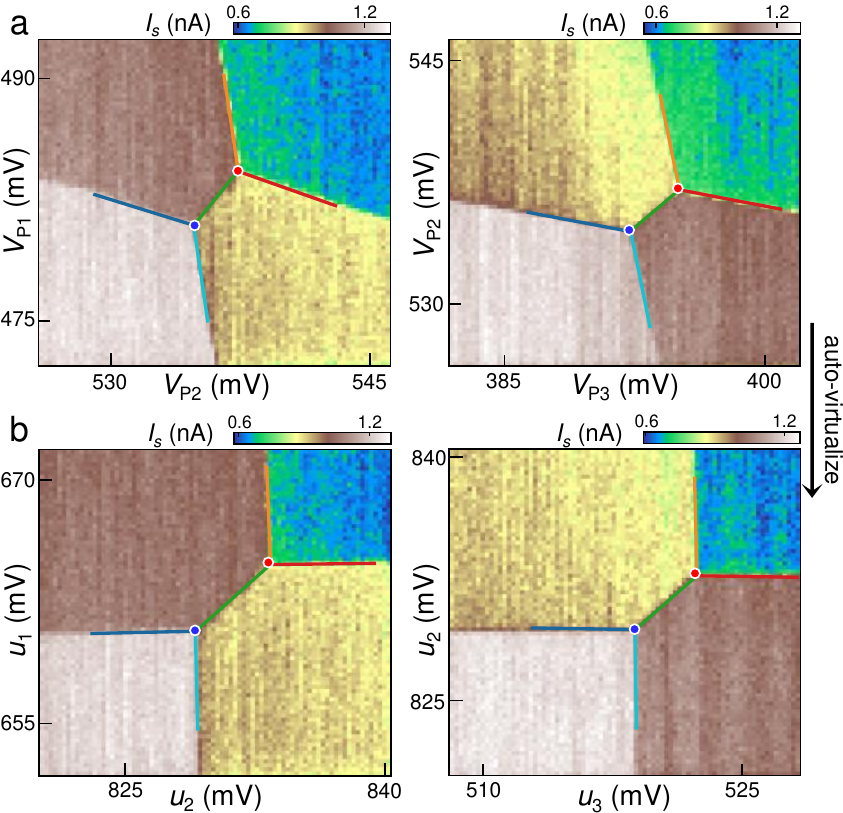}%
\caption{\label{Fig2} (a) Initial pairwise charge stability diagrams for a TQD measured without cross-capacitance corrections: $G_{guess}=I$. (b) Charge stability diagrams measured after auto-virtualizing the plunger gates. The analysis routine took three iterations to converge on an ideal correction matrix.}%
\end{figure}

We begin the automated tuning procedures once the three dots are manually tuned into the single electron regime. In order to establish control over the ground state charge configuration of the TQD, we use the image analysis procedure to automatically compute virtual gates for the device\cite{Hensgens2017,Botzem2018,Volk2019,Mills2019}. As described in detail in Ref.~\onlinecite{Mills2019}, virtual gate voltages $\bm{u}$ are related to plunger gate voltages $\bm{V_P}$ by $\bm{u}=\bm{GV_P}$, where $\bm{G}$ is a dimensionless lever arm matrix related to the capacitance matrix. The virtual gates are implemented in software by inverting $\bm{G}$ and computing the linear combination of voltages $\bm{V_P}$ that compensate for cross-capacitance, allowing the chemical potential of each dot to be individually tuned. The `auto-virtualization' routine begins by measuring pairwise charge stability diagrams centered on the (1,0)-(0,1) interdot charge transitions for both DQD pairs of the TQD in a gate space defined by an initial matrix $\bm{G_{guess}}$. For the first set of measurements, we make no assumptions about $\bm{G}$ and set $\bm{G_{guess}}=\bm{I}$, which corresponds to a normal plunger gate sweep (i.e.\ $\bm{u}=\bm{V_P}$).

Figure 2a shows pairwise charge stability diagrams acquired for dots 1 and 2 (left panel), and dots 2 and 3 (right panel). The chemical potential of the dot not involved in the scan, e.g. dot 3(1) in the left(right) panel of Fig.~2a, is moved above the Fermi level to prevent extra charge transition lines from complicating the auto-fit routine. These data sets are analyzed using the image analysis procedure outlined above, with the fitting results overlaid on the original data in Fig.~2 for clarity. The fitting routine again accurately locates both interdot charge transitions. The slopes of the charge transitions and the locations of the triple points determined from the image analysis routine are used to compute a correction to the capacitance matrix $\bm{G_C}$\cite{Mills2019}. The ideal correction matrix $\bm{G}$ can be expressed as $\bm{G} = \bm{G_C} \cdot \bm{G_{guess}}$. 

We verify that we have a good estimate for $\bm{G}$ by performing another measurement of the charge stability diagrams in the re-defined virtual gate space. When $\bm{G_C}$ is sufficiently close to identity no further corrections need to be made. We quantify convergence by evaluating the diagonal and off-diagonal elements of $\bm{G_C}$. If the diagonal elements are within the range $1\pm 0.1$ and the off-diagonal elements are less than 0.03, we consider $\bm{G}$ converged and exit the routine without applying anymore corrections. These tolerances are primarily influenced by scan fidelity. In this experiment, we scan with a resolution of 0.25~mV per pixel. An error of one pixel on $r_n$ endpoint placement at this fidelity results in an approximate error of $\pm0.03$ in the off-diagonal elements. Similarly, the tolerance for the diagonal elements of $\bm{G_C}$ allows for triple point placement errors of just one or two pixels. In general, taking the data with increased scan fidelity would allow the user to specify tighter constraints. For these data sets, the algorithm converged in three iterations generating the correction matrix $\bm{G} = \begin{psmallmatrix}1 & 0.34 & 0\\ 0.19 & 1.22 & 0.22\\ 0 & 0.20 & 1.04  \end{psmallmatrix}$ that describes the virtual gate parameters $u_{1-3}$ swept in Fig.~2b. As desired, $r_1$ and $r_2$ (as well as $r_4$ and $r_5$) are orthogonal for both DQD pairs, which indicates that the cross-capacitances in the device have been nulled out through the establishment of virtual gates. 


With single electron occupancy in each dot and control over the quantum dot chemical potentials well calibrated through the use of virtual gates, it is now important to optimize interdot tunnel couplings. For example, charge shuttling through the array on a timescale of 50~ns requires interdot tunnel couplings of approximately 20 $\mu$eV or more to maintain adiabaticity\cite{Mills2019}. In practice, interdot tunnel couplings can be adjusted through the barrier gate voltages $V_{B2}$ and $V_{B3}$. However, the barrier gates also couple to dot chemical potentials. In order to auto-tune tunnel coupling, we need to be able to change the bias on our barriers and quickly remeasure the charge stability diagram centered on the interdot transition.

\begin{table}[b]
\begin{tabular}{c|c|c}
\textbf{Barrier - $\mathbf{V_{Bn}}$} 	& $\mathbf{V_{B2}}$ 	& $\mathbf{V_{B3}}$\\
\hline
$\Delta u_{1}^{\rm off}/\Delta V_{Bn}$ 		& $-0.204$			& $-$\\
$\Delta u_{2}^{\rm off}/\Delta V_{Bn}$ 		& $-0.079$ 			& $-0.188$\\
$\Delta u_{3}^{\rm off}/\Delta V_{Bn}$ 		& $-$ 				& $-0.156$\\
\end{tabular}
\caption{\label{Table1} Virtual gate offset shifts, $\Delta u_{i}^{\rm off}$, with barrier gate voltage shifts, $\Delta V_{Bn}$. These measured parameters are used to automatically calculate new triple point locations after barrier gate voltages are changed, keeping the interdot charge transition centered in the measurement window during auto-tuning routines.}
\end{table}

We virtualize the barrier gates by accurately adjusting the offsets of the virtual gates $\bm{u}^{\rm off}$ when barrier gates are tuned\cite{Mills2019}. Here we define the offsets $u_{i}^{\rm off}$ as the intercept of the first charge transition line on the $u_i$ axis and all scans are performed relative to these offsets. We measure the offset drift of each virtual gate as a function of the barrier gate voltages in order to establish virtual barrier gates. The calibration is performed by sweeping across the first charge transition line for each virtual gate parameter $u_i$ while also sweeping the neighboring barrier gate voltage. The slopes of the charge transition lines extracted from these scans, $\Delta u_{i}^{\rm off}/\Delta V_{Bn}$, are listed in Table I. With these data, we compute $\Delta \bm{u^{\rm off}}$ when we change barrier gate voltages and update the offsets $\bm{u^{\rm off}}$ using the formula $\bm{u^{\rm off}} = \bm{u^{\rm off}_0} + \Delta \bm{u^{\rm off}}$ before the next scan. The linear relationship between barrier gate voltages and virtual gate offsets allows us to change the bias on the barrier gates by more than 100 mV and then set up the next charge stability diagram scans with the interdot charge transitions centered on the scan window. Due to the relatively small cross-capacitances in the overlapping gate architecture\cite{Zajac2015} we find that we only need to consider nearest neighbor coupling between barriers and virtual gates, as indicated by the blank fields in Table I.

\begin{figure}[t]
\includegraphics{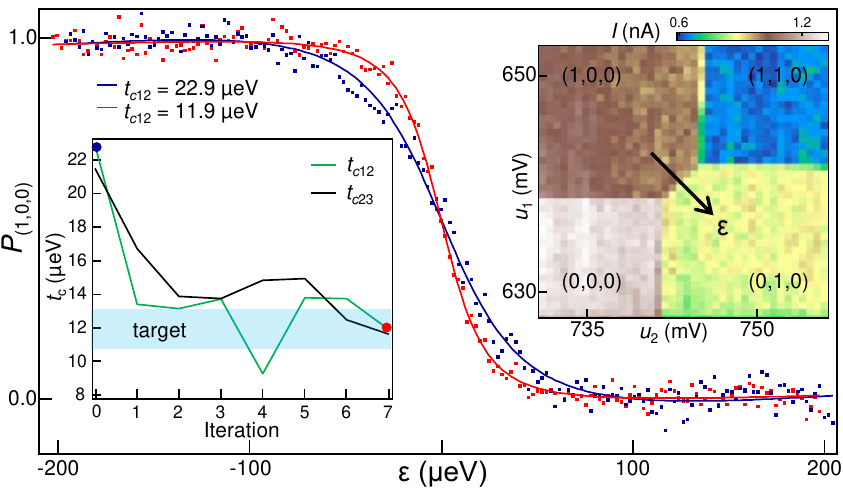}%
\caption{\label{Fig3} Charge sensor response measured as a function of $\epsilon$ along the axis shown in the top right inset and converted to units of dot 1 occupation, $P_{(1,0,0)}$. The bottom left inset shows the measured tunnel couplings at each iteration number. The region of accepted tunnel couplings is shaded in light blue. The blue (red) markers at the start (end) of the $t_{c12}$ curve in the bottom left inset correspond to the blue and red data sets in the main panel.}%
\end{figure}

Once the barriers are virtualized, we can change the interdot tunnel coupling $t_c$ within an auto-tuning algorithm and automatically measure the pairwise charge stability diagrams with the new barrier configuration. We first auto-extract the triple point positions and then measure the charge sensor current $I_s$ as a function of detuning $\epsilon$ (Fig. 3). The resulting curve is fit to extract $t_c$\cite{DiCarlo2004}. We determine the gate lever-arm $\alpha=0.2$ by measuring $I_s(\epsilon)$ at 300 mK, where the thermal energy exceeds the tunnel coupling. With this lever-arm, an electron temperature $T_{el}$ = 55 mK is extracted when $t_c \ll k_BT_{el}$, where $k_B$ is Boltzmann's constant. After the algorithm fits the data and extracts the tunnel couplings, it finds the difference between the target values and the measured values. If all tunnel couplings are within the tolerance of the desired tunnel coupling then the routine exits. Otherwise, the difference is used to determine how much to move the barrier by and the routine repeats, as described below.

We demonstrate simultaneous tuning of the TQD tunnel couplings $t_{c12}$ and $t_{c23}$ by starting at $t_{c12} = 22.9$ $\mu$eV (blue curve in Fig.~3) and $t_{c23} = 21.5$ $\mu$eV and setting a target of 12~$\mu$eV with a tolerance of $\pm$1 $\mu$eV. The inset in Fig.~3 plots the tunnel couplings as a function of iteration number. Here the tuning routine converged after 7 iterations with final tunnel couplings $t_{c12} = 11.9$ $\mu$eV (red curve in Fig.~3) and $t_{c23} = 11.7$~$\mu$eV. We used a constant step parameter $m=3.1$ mV/$\mu$eV to calculate the barrier step size using $\Delta V_{Bn} = m * (t_{target}-t_{c})$ where $\Delta V_{Bn}$ is the change in bias on the barrier $Bn$, $t_{target}$ is the target tunnel coupling and $t_c$ is the current tunnel coupling. We chose $m$ based on preliminary $t_c(V_{Bn})$ measurements and previous auto-tuning runs. Such a simple model worked well enough for this demonstration, but can be further improved by accounting for the non-linearity of $t_c$ as a function of barrier bias.

To accelerate tuning of the interdot tunnel couplings, we set up an automated routine to measure $t_{c12}$ and $t_{c23}$ as a function of barrier gate voltage. We use the data of Fig.~4 to develop a better model of the barrier gate response that is effective in both the low and high $t_c$ regimes. For each measurement, one barrier gate is held at a constant bias while the other barrier gate voltage is stepped between interdot charge transition measurements. As tunnel rates are exponentially related to barrier potential, we use an exponential relation $t_c =t_{c0} + Ae^{(V_{Bn}-V_{Bn0})/\beta}$ to fit the resulting $t_c$ vs $V_{Bn}$ curves and extract an effective ``tunnel coupling lever arm'' $\beta$ for our barrier gates. Here we fix $t_{c0}=k_BT_{el}$ and $A$ = 1 $\mu$eV so that the fit is determined by the horizontal offset $V_{Bn0}$ and $\beta$. The best fits to $t_{c12}$ and $t_{c23}$ are fairly consistent, with $\beta_{12}$ = 25.3~mV and $\beta_{23}$ = 32.9~mV. The difference in horizontal offsets is attributed to imperfections in device fabrication. These results suggest that in future work we can design an algorithm with initial model approximations for all of the barriers in the array and then make active adjustments to the model parameters as measurements are performed during the tuning routines.

\begin{figure}
\includegraphics{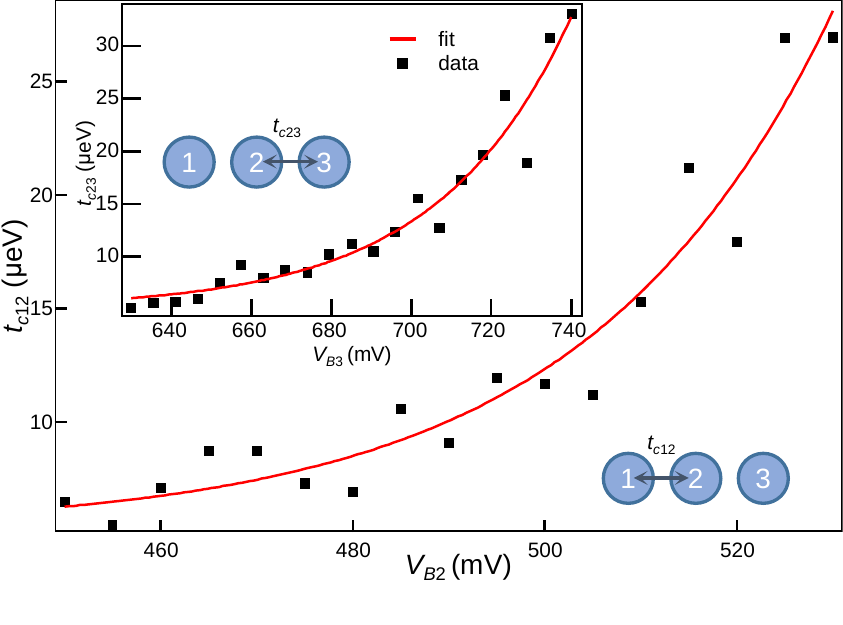}%
\caption{\label{Fig4} Tunnel coupling $t_{c12}$ ($t_{c23}$) measured as a function of $V_{B2}$ ($V_{B3}$). The curves are fit to extract tunnel coupling lever arms $\beta_{12}$ = 25.3 mV and $\beta_{23}$ = 32.9 mV.}%
\end{figure}

In the absence of automation, manually fitting the data and updating device parameters would take a few minutes per double dot pair, in addition to the measurement time. The data analysis demands grow linearly with the number dots in the device, precipitating the need to remove user intervention from the control loop. With the data analysis and tuning adjustments now reduced to a matter of seconds, the main time limitation comes from measurement. The measurement rate can be greatly enhanced by implementing fast sensing approaches\cite{Vink2007,Stehlik2015} which would also allow for higher scan resolution at a lower cost. Improving scan resolution should also reduce the number of iterations required to converge on $\bm{G}$. Finally, using virtual barrier gates in concert with the barrier gate lever arms should greatly reduce the number of iterations required to tune the interdot tunnel couplings in the array.

In summary, we use automated image analysis routines to measure the device capacitance matrix, establish virtual gates, and simultaneously fine-tune both of the interdot tunnel couplings in a TQD.
Tunnel coupling lever arm measurements suggest that more sophisticated auto-tuning algorithms can be developed and also open the door to virtualized tunnel coupling gates in our devices. These automated control routines reduce the amount of effort required to tune up a TQD and may be extended to larger quantum dot arrays\cite{Mills2019}.


%
%

%

\begin{acknowledgments}
We thank Lisa Edge for providing the heterostructure used in these experiments and David Zajac for fabricating the device. Funded by Army Research Office grant W911NF-15-1-0149 and the Gordon and Betty Moore Foundation’s EPiQS Initiative through grant GBMF4535. Devices were fabricated in the Princeton University Quantum Device Nanofabrication Laboratory. The Sandia National Laboratory portion of this work was funded by the Sandia National Laboratory LDRD program and at the Center for Integrated Nanotechnologies, an Office of Science User Facility operated for the U.S. Department of Energy (DOE) Office of Science. Sandia National Laboratories is a multimission laboratory managed and operated by National Technology and Engineering Solutions of Sandia, LLC, a wholly owned subsidiary of Honeywell International, Inc., for the U.S. Department of Energy’s National Nuclear Security Administration under Contract No. DE-NA0003525. This paper describes objective technical results and analysis. Any subjective views or opinions that might be expressed in this paper do not necessarily represent the views of the U.S. Department of Energy or the United States Government.
\end{acknowledgments}

\section*{References}
\bibliography{references_v3}

\end{document}